\documentclass[prl,preprintnumbers,superscriptaddress]{revtex4}

\usepackage{epsf,color,soul,graphicx,amsmath,amssymb}
\usepackage[dvips]{epsfig}
\usepackage{ulem}

\begin{document}

\title{Realization of high-$Q/V$ bichromatic photonic crystal cavities \\     
defined by an effective Aubry-Andr\'e-Harper potential}

\author{A. Simbula}
\affiliation{Department of Physics, University of Pavia, Via Bassi 6, 27100 Pavia, Italy}
\author{M. Schatzl}
\affiliation{Institute of Semiconductor and Solid State Physics, Johannes Kepler University, Altenberger Str. 69, 4040 Linz, Austria}
\author{L. Zagaglia}
\affiliation{Department of Physics, University of Pavia, Via Bassi 6, 27100 Pavia, Italy}
\author{F. Alpeggiani}
\affiliation{Department of Physics, University of Pavia, Via Bassi 6, 27100 Pavia, Italy}
\affiliation{Center for Nanophotonics, FOM Institute AMOLF, Science Park 104, 1098 XG Amsterdam, The Netherlands}
\affiliation{Kavli Institute of Nanoscience, Department of Quantum Nanoscience, Delft University of Technology, Lorentzweg 1, 2628 CJ Delft, The Netherlands}
\author{L. C. Andreani}
\affiliation{Department of Physics, University of Pavia, Via Bassi 6, 27100 Pavia, Italy}
\author{F. Sch\"affler}
\affiliation{Institute of Semiconductor and Solid State Physics, Johannes Kepler University, Altenberger Str. 69, 4040 Linz, Austria}
\author{T. Fromherz}
\affiliation{Institute of Semiconductor and Solid State Physics, Johannes Kepler University, Altenberger Str. 69, 4040 Linz, Austria}
\author{M. Galli}
\affiliation{Department of Physics, University of Pavia, Via Bassi 6, 27100 Pavia, Italy}
\author{D. Gerace}\email{ dario.gerace@unipv.it}
\affiliation{Department of Physics, University of Pavia, Via Bassi 6, 27100 Pavia, Italy}


\begin{abstract}
We report on the design, fabrication and optical {characterization} of bichromatic photonic crystal cavities in thin silicon membranes, with resonances around $1.55$ $\mu$m wavelength. The cavity designs are based on a recently proposed photonic crystal implementation of the Aubry-Andr\'e-Harper bichromatic potential, which relies on the superposition of two one-dimensional lattices with non-integer ratio between the periodicity constants. In photonic crystal  nanocavities, this confinement mechanism is such that optimized figures of merit can be straightforwardly achieved, in particular an ultra-high-Q factor and diffraction-limited mode volume. Several silicon membrane photonic crystal nanocavities with Q-factors in the 1 million range have been {realized}, {as evidenced by resonant scattering}. 
The generality of these designs and their easy implementation and scalability make these results particularly interesting for realizing highly performing photonic nanocavities on different materials platforms and operational wavelengths. 
\end{abstract}


\maketitle

The increasing demand for enhanced optical sensitivity in integrated photonic devices has triggered great progress in the design and realization of photonic crystal (PC) nanocavities, where unprecedented figures of merit have been achieved, such as ultra-high quality factor ($Q$) and diffraction-limited confinement volumes ($V$) in the telecom band \cite{Notomi2010}.  In these nanostructured systems, light-matter interaction can be orders of magnitude larger than in the corresponding bulk medium, making it possible to reach the realm of cavity quantum electrodynamics \cite{Hennessy2007,Faraon2008,Reinhard2012}, with potential applications in prospective quantum photonic technologies \cite{OBrien2009}. With further improvement, nonlinear interactions at the single or few photons level can be expected not only for a strong resonant enhancement of the nonlinearity, i.e. in the presence of a single quantum emitter, but also exploiting the intrinsic higher order material response \cite{Gerace2009,Ferretti2012,Majumdar2013}. More generally, high $Q/V$ PC cavities have potential applications in integrated nonlinear photonics \cite{Notomi2005,Combrie2008}, light emission \cite{Strauf2006,Nomura2010,Ellis2011,Shakoor2013} and sensing \cite{Kwon2008,Jagerska2010}.

The strategies for designing high-$Q/V$ PC slab cavities have strongly benefited from inspirational analogies with condensed matter systems. Following the pioneering demonstration of high-$Q$ point defect cavities in a silicon membrane \cite{Akahane2003}, remarkable $Q/V$ values have been achieved by local modulation of a one-dimensional defect through a shallow trapping potential \cite{Song2005,Kuramochi2006}. 
More recently, a new paradigm in PC cavity engineering has been put forward by means of inverse-design approaches, such as global optimization via genetic algorithms\cite{Minkov2014,Lai2014}. 

Here we report on the design, realization, and optical measurements of high $Q/V$ localized modes in silicon membrane PC cavities, with operational wavelength around $\lambda\sim 1.55$ $\mu$m. These modes originate from the realization of an effective bichromatic potential in a line-defect PC waveguide, which relies on the superposition of two one-dimensional lattices with non-integer ratio between their lattice constants, quantified by the lattice mismatch parameter $\beta = a' / a$, as recently proposed in Ref.~\onlinecite{Alpeggiani2015}. 
This model was originally studied for the Schr\"{o}dinger equation in the context of localization of massive particles in quasi-periodic lattices, and it is commonly known as the Aubry-Andr\'e-Harper (AAH) model \cite{Harper1955,Aubry1980}. The AAH bichromatic potential is known to display a quantum phase transition from extended to localized states as a function of the parameter $\beta$ \cite{Modugno2009,Albert2010}. In particular, for $\beta$ approaching an irrational value, a clear transition between gaussian (also named extended in the literature) to exponentially localized states is predicted, which bears similarities with the Anderson localization in purely disordered systems. Such a transition was first realized in a cold atomic gas \cite{Roati2008}, and later in a photonic lattice of one-dimensional ridge waveguides \cite{Lahini2009}. In the present work, we show {for a selected set of rational $\beta$ values} that such an effective bichromatic potential can be realized in a PC scenario. Remarkably, since we are not in the exponential localization regime of the model, our effective AAH potential naturally leads to a gaussian envelope of the localized modes that is spread over a few lattice sites, which is crucial to achieve ultra-high-$Q$ in PC cavities \cite{Akahane2003,Englund2005}. In fact, we have experimentally measured Q-factors exceeding 1 million in our bichromatic silicon membrane PC cavities, essentially limited by fabrication imperfections since the best theoretically predicted values exceed 1 billion. The theoretical mode volume for these localized modes is estimated in the order of $V\sim (\lambda/n)^3$ from exact numerical simulations, pushing the $Q/V$ values of these cavities among the highest demonstrated so far.

\begin{figure}[t]
\centering
\includegraphics[width=0.8\textwidth]{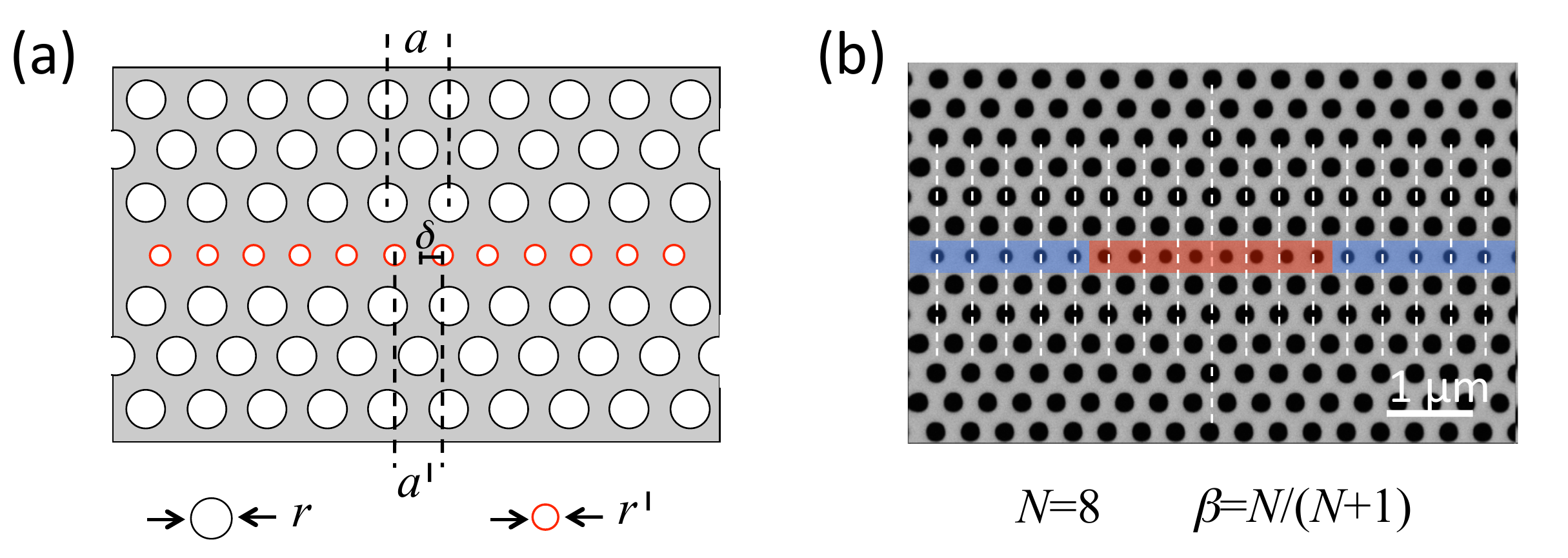}
\caption{(a) Schematic picture of a bichromatic PC {cavity} within a line-defect defined in the triangular lattice of air holes, where $r$ and $a$ represent the radius and lattice constant of the PC triangular lattice, while $r'$ and $a'$ represent radius and lattice constant of the one-dimensional defect lattice, locally shifted by $\delta=a'/2$ with respect to the underlying triangular lattice at the point-defect position. (b) High-resolution {top-view} SEM image showing a typical $N=8$ bichromatic cavity; dashed vertical lines are superimposed to indicate the positions of the underlying triangular lattice points along the line-defect, highlighting the lattice mismatch defined by $a'/a=\beta=N/(N+1)$, where $N$ represents the number of defect holes with reduced radius. The cavity region is evidenced by a red-shaded area, and the one-dimensional mirrors surrounding the cavity along the line-defect are highlighted by blue-shaded  regions.}\label{fig:1}
\end{figure}

The effective bichromatic potential of AAH is realized in the present context by starting from a PC line-defect in a triangular lattice of air holes with lattice constant $a$, the so called W1 waveguide \cite{Joannopoulos_book}. Along the axis of this waveguide, a row of air holes with periodicity $a'$ different from $a$ is introduced, as schematically represented in Fig.~\ref{fig:1}(a). The radius $r'$ of {these} air holes is reduced with respect to the radius {$r$} of the background photonic crystal lattice, which increases the fraction of high-index material. Moreover, the one-dimensional line of reduced air holes is shifted by $\delta=a'/2$ with respect to the original triangular lattice, and a PC point-defect is thus formed. 

As detailed in Ref.~\onlinecite{Alpeggiani2015}, the steady state Maxwell equation for the magnetic field in the PC  structure,
\begin{equation}\label{eq:maxwell}
\nabla \times \left[\frac{1}{\varepsilon(\mathbf{r})}\nabla\times \mathbf{H}(\mathbf{r})\right] = \frac{\omega^2}{c^2} \mathbf{H}(\mathbf{r}) \, ,
\end{equation}
can be approximately recast in a linear eigenvalue problem by expanding the field on a basis of localized states on each site, $\mathbf{H} = \sum c_j \mathbf{H}_j$, where $j$ is an integer labelling the high-index interstitial site along the line-defect with $a'$ periodicity, i.e. corresponding to the positions $x_j =  j a'$  with $j = 0, \pm 1, \pm 2,\ldots$. With this site labelling, $j=0$ corresponds to the center of the point-defect that will constitute the bichromatic cavity. 
For this one-dimensional quasi-periodic lattice, Eq.~\ref{eq:maxwell} can be approximately expressed as
\begin{equation}\label{eq:model}
[\omega^2_0 + \Delta \cos(2\pi \beta j)] c_j - J [c_{j-1} + c_{j+1}] = \omega^2 c_j  \, ,
\end{equation}
where $\omega_0$ is the on-site frequency of the unperturbed lattice, $J$ is the tunnel coupling arising from the overlap of basis functions localized in neighboring interstitial sites (to be meant in a sort of tight-binding approximation, as in Ref.~\onlinecite{Modugno2009}), and $\Delta$ is the on-site amplitude of the modulation potential. The latter can be approximately recovered graphically, i.e. by comparing the dispersion $\omega^2$ vs. $k$ for two {W1} waveguides  filled {by a} row of {air holes with} reduced {radius} $r'$, {either} placed in the original positions of the underlying triangular lattice or shifted by $a/2$ with respect to it \cite{Alpeggiani2015}. 
With this identification of parameters, Eq.~\ref{eq:model} exactly corresponds to the AAH model \cite{Modugno2009,Albert2010,Roati2008,Lahini2009}, for which a quantum phase transition from gaussian to exponential localization regimes is predicted on varying $\Delta/J$, and depending on the degree of commensurability of $\beta$ \cite{footnote}. The phase transitions exactly occurs at the critical ratio $\Delta/J=2$ for irrational $\beta$, and at larger $\Delta/J$ for most rational values of $\beta$ \cite{AnnMath1999}, such as the ones considered in the rest of this  work.

For the realization of the bichromatic PC cavities, we have chosen to avoid the formation of replicas in the
one-dimensional AAH potential. This is achieved by modifying the periodicity of an even number ($N$) of defect holes  within the one-dimensional waveguide channel, which allows to keep the cavity symmetric with respect to the center of the defect. The modified lattice constant for the $N$ holes is chosen to be $\beta = N/(N + 1)$, such that the lattice points located at positions $\pm N a / 2 = (N + 1) a' / 2$ simulateneously belong to both lattices, ensuring a smooth transition between the
cavity defect and the one-dimensional waveguide. The cavity is thus formed by $N$ (even number) defect holes of reduced radius $r'<r$ spaced by $a'$, and then continued by holes with radius $r'$ at positions $na$ along the W1 waveguide channel (for every integer $|n| \ge N/2$). {As an} example, a scanning electron microscope (SEM) top-view image of an implemented cavity with $N=8$  is {shown} in Fig.~\ref{fig:1}(b). The lattice constant of the device shown is $a={400}$ nm, and the defect lattice is given by the periodicity $a'=Na/(N+1)={356}$ nm, while $r={115}$ nm, and $r'={82}$ {nm}. We have realized several devices on the same chip, with $N$ ranging from $4$ to $24$ (only even values), corresponding to an effective AAH model realized for a selected set of (rational) lattice mismatch values, $\beta$, ranging from $0.8$ to $0.96$.

\begin{figure}[t]
\centering
\includegraphics[width=0.9\textwidth]{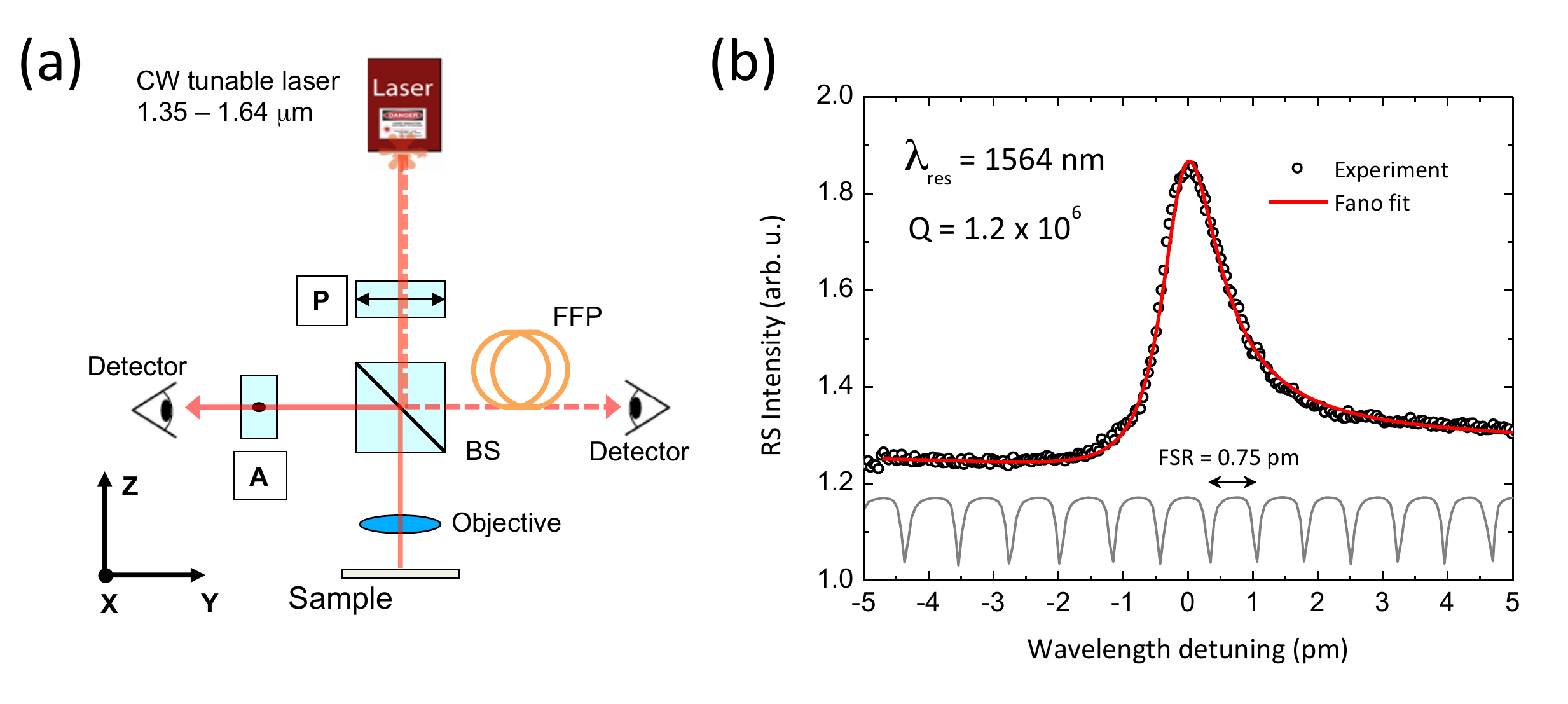}
\caption{(a) Schematic illustration of the resonant scattering set-up employed to perform high-resolution spectral detection of the PC cavity signal in cross-polarization: P is a linear polarizer, A is an orthogonally polarized analyzer, BS an optical beam splitter, and FFP is a fiber-Fabry-P\'{e}rot interferometer.  (b) A typical resonant scattering spectrum measured on one of the high-$Q$ bichromatic cavity samples (points). The corresponding fit with a Fano-like lineshape  is shown with full line, resulting in the quoted Q-factor and telecom resonant wavelength; the Fabry-P\'{e}rot spectrum (FSR) used for wavelength calibration is displayed at the bottom of the plot (grey line).}\label{fig:2}
\end{figure}

The fabrication of the bichromatic PC cavities was performed on standard silicon-on-insulator (SOI) wafers commercially available from SOITEC, with a 220 nm-thick Si core layer and 2 $\mu$m of SiO$_2$ {buried} oxide. Photonic patterns are defined {on $50\times 50\,\mu\text{m}^2$ exposure fields} by electron-beam lithography {using an AR-P679.04 PMMA resist and an AR 600-56 developer, both obtained from ``Allresist''. The pattern was transferred into the Si layer using a SF$_6$/O$_2$ based cryo-process in a Oxford P100 inductive coupled plasma reactive ion etcher. Subsequently, the buried SiO$_2$ layer was removed by immersing the sample for 2 minutes into a 40\% HF solution, resulting in suspended PC membranes. During the HF etch, the membrane top surface was protected by the PMMA resist mask to prevent roughening of the Si surface. Finally, the PMMA mask was removed in a O$_2$ plasma etcher and the sample was rinsed in acetone and methanol prior to a final 10-second dip in a 2\% HF solution.} In the following, we will concentrate on the results obtained for the set of devices with lattice constant $a=430$ nm, and nominal $r=120$ nm, $r'=90$ nm. {Based on our}  SEM {data}, we estimate {$r$ and $r'$ to agree within $\pm3\%$ with the nominal parameters}.

\begin{figure}[t]
\centering
\includegraphics[width=0.95\textwidth]{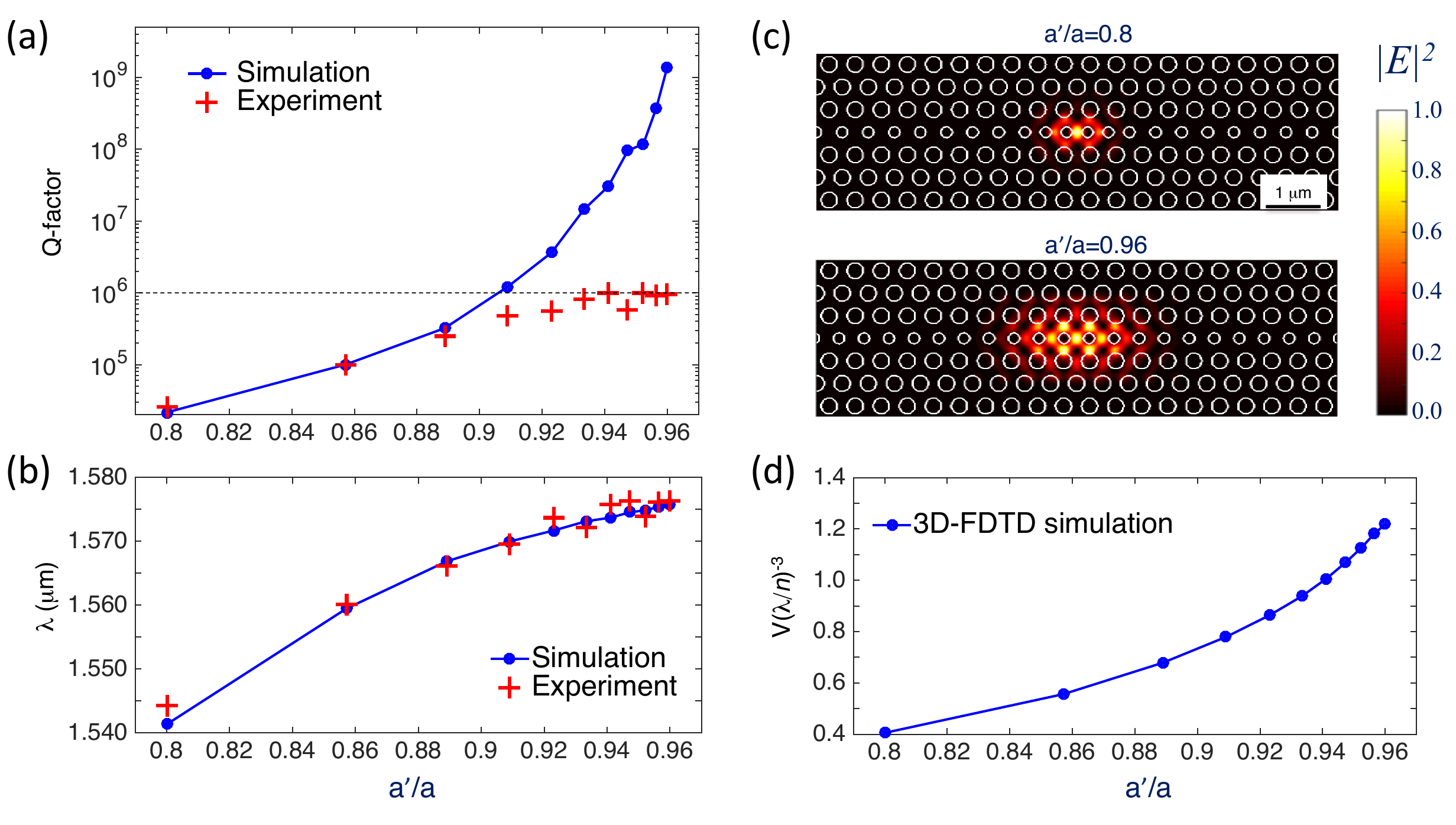}
\caption{Experimental and simulated results for bichromatic cavities with PC lattice constant $a=430$~nm and radii  $r=120$~nm, $r'=90$~nm:  
(a) Quality factor of the bichromatic PC cavities as a function of the AAH lattice mismatch parameter, $\beta=a'/a$, (b) corresponding resonant wavelength as measured from the resonant scattering spectra. The 3D-FDTD calculated $Q$ and $\lambda$ are shown for a direct comparison, and the dashed line indicates the $Q=1$ million threshold. (c) Simulated electric field intensity profiles for the shortest and longest cavities ($N=4$  and $N=24$, respectively) investigated in this work, represented in a color scale contour plot; in both cases, the intensities have been normalized to the peak values at the cavity antinodes. (d) Calculated cavity mode volumes (see Eq.~\ref{eq:volume}) as a function of $\beta$.}\label{fig:3}
\end{figure}

The fabricated bichromatic PC cavities were experimentally tested by a cross-polarized resonant scattering technique \cite{Galli2009}. This technique is particularly suited for measuring PC cavities with ultra-high $Q$-factors, since it does not rely on evanescent coupling (such as, e.g., through an integrated photonic crystal waveguide or an external fiber taper) \cite{Meccuccio2005}, thus avoiding any loading effect that may reduce the intrinsic value of the measured $Q_{exp}$.  A scheme of the optical setup is illustrated in Fig.~\ref{fig:2}(a). Here, the linearly polarized light from a tunable laser with a  $\Delta\nu<10$ MHz linewidth is used to resonantly excite the PC cavity mode from the top through a NA=0.8 microscope objective. The back-scattered light from the cavity is then collected by the same objective and sent to the analyzer. A careful calibration of the laser detuning wavelength is of utmost importance to achieve an accurate estimation of the ultra-high $Q$, which was obtained by the simultaneous measurement of a reference spectrum from a 2 meters long fiber-Fabry-P\'{e}rot (FFP) interferometer with a 0.75 pm free spectral range.
In Fig.~\ref{fig:2}(b) we also show a typical spectral response detected from one of our highest $Q$ bichromatic cavities, displaying a remarkable experimental $Q\sim 1.2 \times 10^6$ at the resonant wavelength $\lambda\sim 1.56$ $\mu$m.

The experimental Q-factors and cavity mode wavelengths are extracted from the resonant scattering spectra and shown in Fig.~\ref{fig:3}(a-b) for a series of bichromatic PC cavities as a function of $\beta$. The experimental data are compared to full three-dimensional finite difference time domain (3D-FDTD) simulations \cite{lumerical} of the measured devices, where all the nominal fabrication parameters have been assumed, and $n=3.5$ is taken as the dispersionless refractive index of silicon in the wavelength range {of interest}. Very good  agreement is found up to $\beta\sim 0.889$ (i.e., $N=8$), both in terms of Q-factor and resonant wavelength. Then, while the measured wavelengths are very well matched by numerical simulations, Q-factors start to deviate and roughly saturate to values on the order of $Q_{exp}\sim 10^6$. The theoretical $Q$ values increase up to values well exceeding $Q_{theo}=10^9$ at $\beta=0.96$ ($N=24$), as already reported \cite{Alpeggiani2015}. 
{The extremely large $Q_{theo}$-values} can be {understood} by fitting the AAH model parameters {suitable} for the present devices, which we have determined as $\Delta=0.00805(a/2\pi c)^2$ and $J=0.00715(a/2\pi c)^2$, respectively. Hence, the value $\Delta/J\sim 1.125$ automatically yields a gaussian localization regime for $\beta<1$, the critical value for the transition to exponential  localization of the field being at least $\Delta/J=2$ \cite{Modugno2009,Aubry1980,Albert2010,AnnMath1999}. As it is accepted in the PC cavity literature, a perfectly gaussian envelope is a favorable condition to achieve ultra-high Q-factors in PC slab cavities \cite{Akahane2003,Englund2005}. In the bichromatic PC cavity designs such ultra-high $Q$ condition is met quite straightforwardly, depending on a single design parameter ($\beta$), and without the need for extensive numerical simulations.
Regarding the deviation of the measured Q-factors from the theoretical ones that appears for $\beta>0.9$, this is mainly attributed to the role of fabrication imperfections: the experimental values can be expressed as $Q_{exp}^{-1}=Q_{theo}^{-1}+Q_{ext}^{-1}$, where $Q_{ext}$ is the Q-factor determined by extrinsic losses introduced in the fabrication process, such as {disorder and roughness of the lower PC membrane surface exposed to the 40\% HF etch during SiO$_2$ removal as well as } material absorption \cite{Gerace2005,Portalupi2011}. In the present samples, it can be inferred that the limiting Q-factor induced by extrinsic losses is on the order of $Q_{ext}\sim 10^6$, which makes $Q_{exp}$ saturate at such values even when $Q_{theo}$ is orders of magnitude larger.  A detailed study on the role of disorder in these bichromatic PC cavities would certainly be interesting but it is beyond the scope of the present manuscript.

\begin{figure}[t]
\centering
\includegraphics[width=0.8\textwidth]{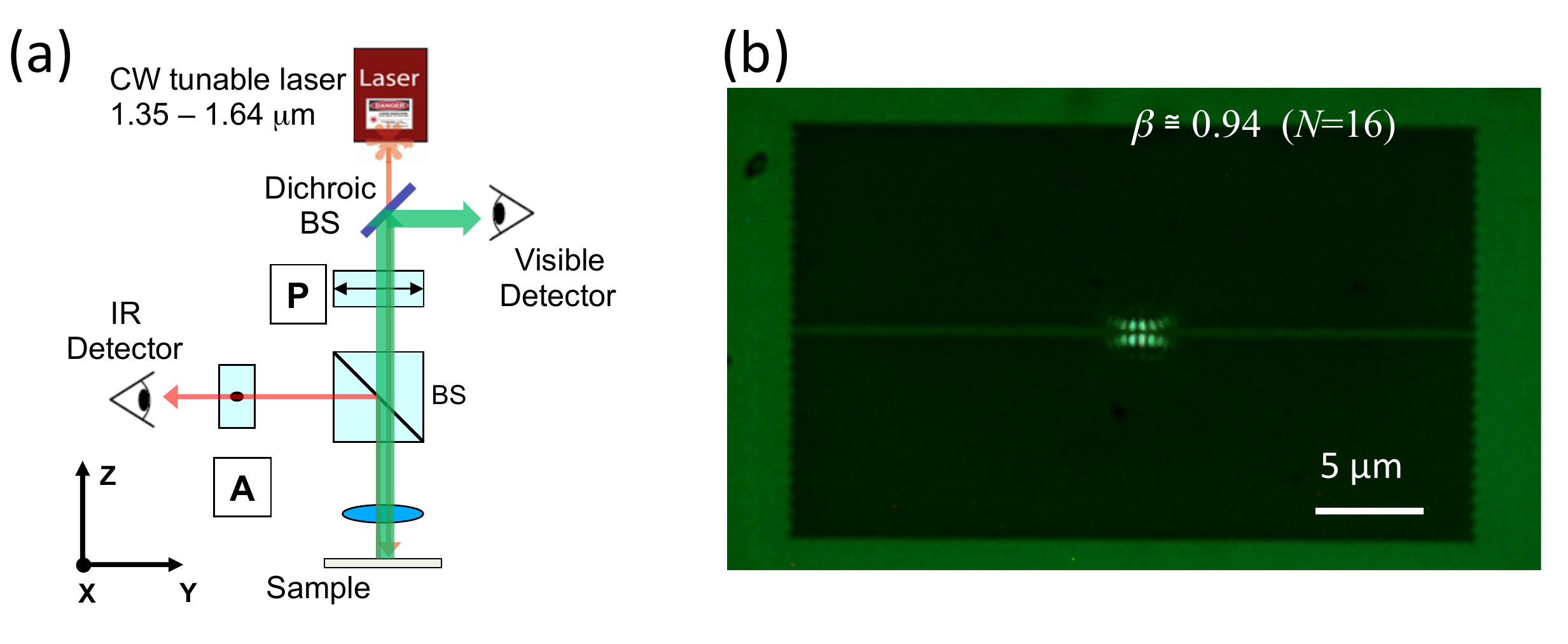}
\caption{(a) Schematic representation of the modified set-up allowing to image (in the visible) the third-harmonic signal emitted from the bichromatic PC cavity  excited at the resonant wavelength (in the near-infrared), and (b) typical green light emission from the third harmonic of one of the fabricated samples (in this case, with $N=16$), showing the localized nature of the bichromatic cavity mode in a diffraction-limited spot.}\label{fig:4}
\end{figure}

The mode volume of PC cavities can be directly calculated from the steady state electric field profile of the localized modes simulated by 3D-FDTD. {As examples, we explicitly show in} Fig.~\ref{fig:3}(c) {the mode profiles} for the smallest ($\beta=0.8$, or $N=4$) and the largest ($\beta= 0.96$, or $N=24$) {bichromatic} cavity considered in this work, respectively. 
Clearly, the spatial extension of the mode {increases with} $\beta$. To give a quantitative estimate, we calculate the mode volume according to the usual cavity quantum electrodynamics definition
\begin{equation}\label{eq:volume}
V = \int \mathrm{d}\mathbf{r}\,\frac{\varepsilon(\mathbf{r})  |E(\mathbf{r})|^2} {\mathrm{max}\{\varepsilon(\mathbf{r})|E(\mathbf{r})|^2\}} \, ,
\end{equation}
which is relevant for a single dipole emitter located at the electric field intensity maximum (i.e., in the cavity center in this case) \cite{Notomi2010}. The mode volumes calculated for the whole series of {experimentally characterized cavities presented} in Figs.~\ref{fig:3}(a-b) are shown in Fig.~\ref{fig:3}(d), normalized to the corresponding $(\lambda/n)^3$, where $\lambda$ is the resonant wavelength of each bichromatic PC cavity and $n=3.5$  is the Si refractive index assumed in the simulations. As it can be seen, the mode volume remains of the order of the diffraction limit, $(\lambda/n)^3$, even for the largest bichromatic PC cavities. In particular, for the $\beta=0.96$ cavity the theoretical $Q_{theo}/V\sim 1.5\times 10^9 (\lambda/n)^{-3}$ ratio is among the largest reported for 2D PC slab cavities and can be even larger \cite{Alpeggiani2015}, competing in particular with the highest $Q_{theo}/V\sim 2.5\times 10^9 (\lambda/n)^{-3}$ design of an index-modulated cavity in a PC waveguide \cite{Notomi2008}. If we relate the theoretical mode volumes to the measured Q-factors saturating at $Q_{exp} \sim 10^6$ due to {surface roughening and} disorder-induced losses, we see that the best compromise is obtained for the devices corresponding to $\beta\simeq 0.941$ ($N=16$) and $\beta\simeq 0.952$ ($N=20$), which display a remarkable $Q_{exp}/V\sim 10^6 (\lambda/n)^{-3}$, i.e.  among the largest ever reported for PC slab cavities \cite{Lai2014}.

Finally, we report in Fig.~\ref{fig:4} the results of a nonlinear optical experiment performed on a slightly modified setup, schematically illustrated in Fig.~\ref{fig:4}(a), which allows to record the higher-order harmonic emission from the sample. The selected PC cavity device is resonantly excited with a continuous wave laser at the localized mode wavelength, and nonlinear light emission is collected and imaged with a detector in the visible range (in this case, a commercially available Nikon camera). 
As an example, we show in Fig.~\ref{fig:4}(b) the green light emission from a $\beta\simeq 0.94$ ($N=16$) cavity, excited at a resonant wavelength of $\lambda=1563$ nm, and thus emitting in the third harmonic at $\lambda_{TH}=521$ nm, which gives the green signal. We notice that the third harmonic is generated by the bulk $\chi^{(3)}$ susceptibility of silicon, which is strongly enhanced by the confinement of the cavity mode in a $V\sim (\lambda/n)^{3}$ effective volume, and it thus allows to detect the nonlinear signal even under cw excitation, as already shown previously \cite{Galli2010}. To the purposes of the present work, this image gives a direct evidence of the gaussian localization of the resonant modes in the bichromatic PC cavities.

{In summary, we have shown the first realization of a novel type of photonic crystal slab cavities, based on the analogy with the Aubry-Andr\'e-Harper model in condensed matter physics. The gaussian confined modes arise from the lattice mismatch between a linear photonic crystal waveguide and a defect row of reduced radius holes inserted within the waveguide channel. The experimentally measured {$Q$-factor} values are among the largest reported for photonic crystal slab cavities, only limited to values in the 1 million range by fabrication imperfections. The corresponding theoretical Q-factors can easily exceed 1 billion, thus potentially making these cavities the ones with the largest $Q/V$ ratio ever designed. Since they do not require any sophisticated design strategy or time-consuming numerical simulations, but the tuning of a single parameter like the lattice mismatch, these devices are particularly promising for a fast realization of highly performing photonic crystal cavities for a number of applications in integrated photonics, nonlinear optics, light emission, and sensing.



\begin{thebibliography}{10}
\bibitem{footnote}
In comparing the photonic AAH model to the original one, notice that the solutions of the eigenvalue equation derived from Eq.~\ref{eq:maxwell} are frequencies squared, and not the eigenenergies. Hence, in Eq.~\ref{eq:model} also $J$ and $\Delta$ have dimensions of $\omega^2$.
\end{thebibliography}
\end{document}